# On the infinite gradient-flow for the domain-wall formulation of chiral lattice gauge theories

Taichi Ago[a] and Yoshio Kikukawa[b]

[a]*Department of Physics, University of Tokyo, Tokyo 113-0033, Japan*
[b]*Institute of Physics, University of Tokyo, Tokyo 153-8092, Japan*

 *E-mail:* ago@hep-th.phys.s.u-tokyo.ac.jp, kikukawa@hep1.c.u-tokyo.ac.jp

ABSTRACT:
We examine the proposal by Grabowska and Kaplan (GK) to use the infinite gradient flow in the domain-wall formulation of chiral lattice gauge theories. We consider the case of Abelian theories in detail, for which Lüscher's exact gauge-invariant formulation is known, and we relate GK's formulation to Lüscher's one. The gradient flow can be formulated for the admissible U(1) link fields so that it preserves their topological charges. GK's effective action turns out to be equal to the sum of Lüscher's gauge-invariant effective actions for the target Weyl fermions and the mirror "fluffy" fermions, plus the so-called measure-term integrated along the infinite gradient flow. The measure-term current is originally a local(analytic) and gauge-invariant functional of the admissible link field, given as a solution to the local cohomology problem. However, with the infinite gradient flow, it gives rise to non-local(non-analytic) vertex functions which are not suppressed exponentially at large distance. The "fluffy" fermions remain as a source of non-local contribution, which couple yet to the Wilson-line and magnetic-flux degrees of freedom of the dynamical link field.

# Contents

## 1 Introduction

In the recent proposal by Grabowska and Kaplan[1, 3] of a non-perturbative regularization for chiral gauge theories, the authors consider the original (periodic) domain wall fermion by Kaplan[4, 5], but coupled to the "five-dimensional" link field which is obtained by the gradient flow[6–9] from the dynamical four-dimensional link field at the target wall toward the mirror wall. This choice of the "five-dimensional" link field makes possible a chiral gauge-coupling for the target and mirror walls, while keeping the system four-dimensional and gauge-invariant.

However, there is some doubt about locality in this formulation. In the weak gauge-coupling limit of the topologically trivial sector, the condition that the form factor of the mirror-modes is soft enough to suppress the transverse gauge-coupling is given by $\sum_\mu 4\sin^2(p_\mu/2)\,t \gg 1$ for all possible momenta $p_\mu \in [-\pi, \pi]$ above a certain IR cutoff $\mu$, where $t$ is the flow-time of gradient flow. If one assumes $|p_\mu| \geq \mu \equiv \pi/L$ where $L$ is the linear extent of a lattice, the condition reads $t \gg L^2/\pi^2$ [7–9]. This apparently contradicts the other condition $1 \ll \sqrt{8t} \ll L$ for that a local composite operator of the flowed five-dimensional link field is local w.r.t. the original four-dimensional link field. This implies that the imaginary part of the effective action, which is written with the local operators of the five-dimensional link field, can contain the *non-local* operators w.r.t. the dynamical four-dimensional link field. This is one reason why the gauge-invariance is maintained in this formulation even when any anomalous set of chiral-modes appear in the target wall. The fate of the possible non-local terms in anomaly-free cases is not fully clarified yet.

In this paper, we examine the proposal by Grabowska and Kaplan (GK) to use the (infinite) gradient flow in the domain-wall formulation of chiral lattice gauge theories.[1] We consider the case of Abelian theories in detail, for which an exact gauge-invariant formulation has been obtained[13] based on the Ginsparg-Wilson relation[14, 15]/the overlap Dirac operator[16–20] and its relation to domain wall formulation has been clarified[21]. Based on these known results, we relate GK's formulation to the exact gauge-invariant one and examine the locality property of the former in relation to the latter. The gradient flow can be formulated for the admissible U(1) link fields so that it preserves their topological charges. GK's effective action turns out to be equal to the sum of Lüscher's gauge-invariant effective actions for the target Weyl fermions and the mirror "fluffy" fermions, plus the so-called measure-term integrated along the infinite gradient flow. The measure-term current is originally a local(analytic) and gauge-invariant functional of the admissible link field, given as a solution to the local cohomology problem. However, with the infinite gradient flow, it gives rise to non-local(non-analytic) vertex functions $\Gamma^{(k)}(x_1, \ldots, x_k)$ which are only suppressed at large distance as $r^{-(5k-4)}$ where $r = \sum_{i=2}^{k} |x_i - x_1|_1$. The "fluffy" fermions remain as a source of non-local contribution, which couple yet to the Wilson-line and magnetic-flux degrees of freedom of the dynamical link field.

This paper is organized as follows. In section 2, we first review briefly the construction of U(1) chiral lattice gauge theories with exact gauge invariance based on the overlap Dirac operator/the Ginsparg-Wilson relation [13], and its relation to the domain wall fermion [21]. In section 3, we formulate the gradient flow for the admissible U(1) link fields. In section 4, we implement the proposal by Grabowska and Kaplan using the simplified domain wall fermion[5], and clarify its relation to Luscher's gauge-invariant formulation. In section 5, we discuss the issue of locality. In section 6, we conclude with some discussions.

## 2 Abelian chiral gauge theories on the lattice with exact gauge invariance

In this section, we first review briefly the construction of U(1) chiral lattice gauge theories with exact gauge invariance based on the overlap Dirac operator/the Ginsparg-Wilson

---

[1] We do not consider here the so-called chiral solution of the Ginsparg-Wilson relation[2] and the related issues[10–12]. As far as the authors understand, the chiral solution is resulted by a wrong use of the subtraction scheme which is only applicable to vector-like theories, where the sizable direct coupling among the target wall with $U(x,\mu)$ and the mirror wall with $U_\star(x,\mu)$ is introduced through the anti-periodic boundary condition for the Pauli-Villars field. In fact, as shown in appendix A for detail, even for the gradual flow, one obtains the same chiral solution irrespective of the flow types, the gradual flow or the sudden flow, as long as one uses the wrong subtraction scheme. See section 2 about the correct use of the Pauli-Villars spinor-boson with the anti-periodic boundary condition for chiral gauge theories.

relation [13], and its implementation based on the domain wall fermion [21].

We consider four-dimensional U(1) gauge theories where the gauge field couples to $N$ left-handed Weyl fermions with charges $q_\alpha$ satisfying the anomaly cancellation condition[2],

$$\sum_{\alpha=1}^{N}(q_\alpha)^3 = 0. \tag{2.2}$$

We assume the four-dimensional lattice of the finite size $L$ and choose lattice units,

$$\Gamma = \left\{ x = (x_1, x_2, x_3, x_4) \in \mathbb{Z}^4 \mid 0 \leq x_\mu < L \, (\mu = 1, 2, 3, 4) \right\}, \tag{2.3}$$

and adopt the periodic boundary condition for both boson fields and fermion fields in the four-dimensional lattice[3].

## 2.1 Gauge fields

As for the U(1) gauge fields, we require the so-called admissibility condition:

$$|F_{\mu\nu}(x)| < \epsilon \qquad \text{for all } x, \mu, \nu, \tag{2.4}$$

where the field tensor $F_{\mu\nu}(x)$ is defined from the plaquette variables,

$$F_{\mu\nu}(x) = \frac{1}{i}\ln P_{\mu\nu}(x), \quad -\pi < F_{\mu\nu}(x) \leq \pi, \tag{2.5}$$

$$P_{\mu\nu}(x) = U(x,\mu)U(x+\hat{\mu},\nu)U(x+\hat{\nu},\mu)^{-1}U(x,\nu)^{-1}, \tag{2.6}$$

and $\epsilon$ is a fixed number in the range $0 < \epsilon < \pi$. This condition ensures that the overlap Dirac operator[16, 17] is a smooth and local function of the gauge field if $|q_\alpha|\epsilon < 1/30$ for all $\alpha$ [22].[4] The admissibility condition may be imposed dynamically by choosing the following action,

$$S_G = \frac{1}{4g_0^2}\sum_{x\in\Gamma}\sum_{\mu,\nu} L_{\mu\nu}(x), \tag{2.7}$$

where

$$L_{\mu\nu}(x) = \begin{cases} [F_{\mu\nu}(x)]^2 \left\{1 - [F_{\mu\nu}(x)]^2/\epsilon^2\right\}^{-1} & \text{if } |F_{\mu\nu}(x)| < \epsilon, \\ \infty & \text{otherwise.} \end{cases} \tag{2.8}$$

The admissible U(1) gauge fields can be classified by the magnetic fluxes,

$$m_{\mu\nu} = \frac{1}{2\pi}\sum_{s,t=0}^{L-1} F_{\mu\nu}(x+s\hat{\mu}+t\hat{\nu}), \tag{2.9}$$

---

[2] We also consider two-dimensional U(1) gauge theories where the gauge field couples to $N$ left-handed Weyl fermions with charges $q_\alpha$ and $N'$ right-handed Weyl fermions with charges $q'_{\alpha'}$ satisfying the anomaly cancellation condition,

$$\sum_{\alpha=1}^{N}(q_\alpha)^2 - \sum_{\alpha'=1}^{N'}(q'_{\alpha'})^2 = 0. \tag{2.1}$$

[3] Here we adopt the periodic boundary conditions as in [13]. We can also adopt other types of periodic boundary condition, but the result is unaffected in the infinite volume limit.

[4] In two dimensions, the condition reads $|q_\alpha|\epsilon < 2/5$ for all $\alpha$ and $|q'_{\alpha'}|\epsilon < 2/5$ for all $\alpha'$

which are integers independent of $x$. We denote the space of the admissible gauge fields with a given magnetic flux $m_{\mu\nu}$ by $\mathfrak{U}[m]$. As a reference point in the given topological sector $\mathfrak{U}[m]$, one may introduce the gauge field which has the constant field tensor equal to $2\pi m_{\mu\nu}/L^2 (< \epsilon)$ by

$$V_{[m]}(x,\mu) = e^{-\frac{2\pi i}{L^2}[L\delta_{\tilde{x}_\mu, L-1}\sum_{\nu>\mu}m_{\mu\nu}\tilde{x}_\nu + \sum_{\nu<\mu}m_{\mu\nu}\tilde{x}_\nu]} \quad (\tilde{x}_\mu = x_\mu \bmod L). \qquad (2.10)$$

Then any admissible U(1) gauge field in $\mathfrak{U}[m]$ may be expressed as

$$U(x,\mu) = e^{iA^T_\mu(x)} \, U_{[w]}(x,\mu) \, \Lambda(x) \, \Lambda(x+\hat{\mu})^{-1} \, V_{[m]}(x,\mu), \qquad (2.11)$$

where $A^T_\mu(x)$ is the transverse vector potential uniquely determined by

$$\partial^*_\mu A^T_\mu(x) = 0, \qquad \sum_{x\in\Gamma} A^T_\mu(x) = 0, \qquad (2.12)$$

$$\partial_\mu A^T_\nu(x) - \partial_\nu A^T_\mu(x) + 2\pi m_{\mu\nu}/L^2 = F_{\mu\nu}(x), \qquad (2.13)$$

$U_{[w]}(x,\mu)$ represents the degrees of freedom of the Wilson lines,

$$U_{[w]}(x,\mu) = \begin{cases} w_\mu & \text{if } x_\mu = L-1, \\ 1 & \text{otherwise}, \end{cases} \qquad (2.14)$$

with the phase factor $w_\mu \in U(1)$ and $\Lambda(x)$ is the gauge function satisfying $\Lambda(0) = 1$.

## 2.2 Weyl fields

Weyl fields are introduced based on the overlap Dirac operator which satisfies the Ginsparg-Wilson relation[16, 17]:

$$D = \frac{1}{2}\left(1 + X\frac{1}{\sqrt{X^\dagger X}}\right), \qquad (2.15)$$

$$X = -\gamma_\mu(\nabla_\mu - \nabla^\dagger_\mu) + \frac{1}{2}\nabla^\dagger_\mu\nabla_\mu - m_0, \quad 0 < m_0 < 2, \qquad (2.16)$$

where $\partial_\mu$ ($\nabla_\mu$) are the forward (covariant) difference operators. We first consider the overlap Dirac fields $\psi(x)$ which carry a Dirac index $s = 1, 2, 3, 4$ and a flavor index $\alpha = 1, \cdots, N$. Each component $\psi_\alpha(x)$ couples to the U(1) link fields, $U(x,\mu)^{q_\alpha}$. We assume that the lattice overlap Dirac operators $D$ acting on $\psi(x)$, respectively satisfy the Ginsparg-Wilson relation,

$$\gamma_5 D + D\hat{\gamma}_5 = 0, \qquad \hat{\gamma}_5 \equiv \gamma_5(1 - 2D), \qquad (2.17)$$

and we define the projection operators as

$$P_\pm = \left(\frac{1 \pm \gamma_5}{2}\right), \quad \hat{P}_\pm = \left(\frac{1 \pm \hat{\gamma}_5}{2}\right), \qquad (2.18)$$

The left-handed Weyl fermions can be defined by imposing the constraints,

$$\psi_-(x) = \hat{P}_-\psi(x), \quad \bar{\psi}_-(x) = \bar{\psi}(x)P_+. \qquad (2.19)$$

## 2.3 Gauge-invariant effective action

The gauge-invariant effective action of the overlap Weyl fermions[13] is then given by

$$e^{\Gamma[U]} = \det(\bar{v}_k D v_j), \qquad (2.20)$$

where $\{v_j(x)\}$ and $\{\bar{v}_k(x)\}$ are the orthonormal chiral bases satisfying

$$\hat{P}_- v_j(x) = v_j(x), \quad \bar{v}_k(x) P_+ = \bar{v}_k(x). \qquad (2.21)$$

$\{\bar{v}_k(x)\}$ is a fixed basis independent of the link field and $\{v_j(x)\}$ is specified by the following formulae: for any interpolation of the link fields $U_s(x,\mu)$ ($s \in [0,1]$) within $\mathfrak{U}[m]$ from a reference link field $U_0(x,\mu)$ to the target one $U_1(x,\mu) = U(x,\mu)$,

$$v_j(x) = \begin{cases} Q_1 v_1^0(x) W^{-1} & \text{if } j = 1 \\ Q_1 v_j^0(x) & \text{otherwise} \end{cases} \qquad (2.22)$$

$$\partial_s Q_s = [\partial_s \hat{P}_s, \hat{P}_s] Q_s, \quad Q_0 = 1, \quad \hat{P}_s = \hat{P}_-|_{U=U_s}, \qquad (2.23)$$

$$W = \exp\left\{ \int_0^1 ds \sum_{x \in \Gamma} \sum_\mu \partial_s U_s(x,\mu) U_s(x,\mu)^{-1} j_\mu(x) \right\}. \qquad (2.24)$$

Here, the local current $j_\mu(x)$ is constructed so that it satisfies the following four properties,

1. $j_\mu(x)$ is defined for all admissible gauge fields and depends smoothly on the link variables.

2. $j_\mu(x)$ is gauge-invariant and transforms as an axial vector current under the lattice symmetries.

3. The linear functional $\mathfrak{L}_\eta = \sum_{x \in \Gamma} \eta_\mu(x) j_\mu(x)$ is a solution of the integrability condition

$$\delta_\eta \mathfrak{L}_\zeta - \delta_\zeta \mathfrak{L}_\eta = i \text{Tr}\left\{ \hat{P}_- [\delta_\eta \hat{P}_-, \delta_\zeta \hat{P}_-] \right\} \qquad (2.25)$$

for all periodic variations $\eta_\mu(x)$ and $\zeta_\mu(x)$, where the variation of the link field $U(x,\mu)$ is defined by $\delta_\xi U(x,\mu) = i\xi_\mu(x) U(x,\mu)$ for $\xi_\mu(x) = \eta_\mu(x), \zeta_\mu(x)$.

4. The anomalous conservation law holds:

$$\partial_\mu^* j_\mu(x) = \text{tr}\{Q\gamma_5(1-D)(x,x)\}, \qquad (2.26)$$

where $Q = \text{diag}(q_1, \cdots, q_N)$.

## 2.4 Relation to five-dimensional domain wall fermion

The gauge-invariant effective action formulated by Lüscher, $\Gamma[U]$, is related to the (subtracted) partition function of the domain wall fermion with a local counter term by the following formula [21]:

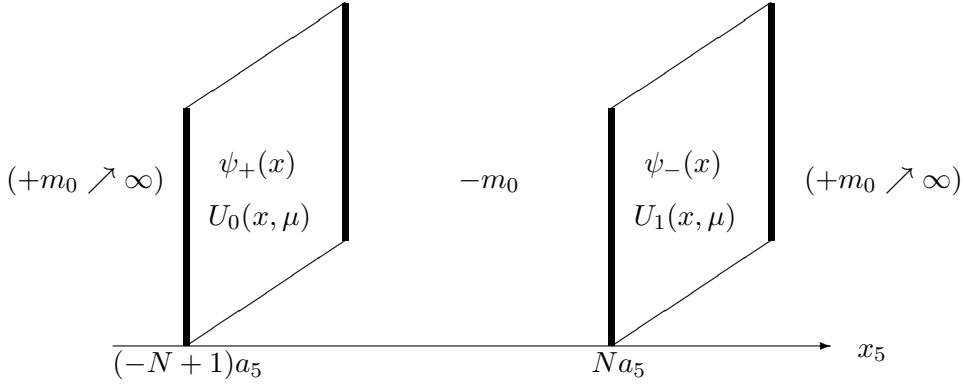

**Figure 1**. Domain wall fermion in the simpler vector-like setup

$$e^{\Gamma[U_1]} e^{\Gamma[U_0]^*} = \det\{1 - P_+ + P_+ D Q_1 D_0^\dagger\} e^{-i \int_0^1 ds \, \mathfrak{L}_{\eta^s}} \tag{2.27}$$

$$= \lim_{a_5 \to 0} \lim_{N \to \infty} \frac{\det (D_{5w} - m_0)|_{\text{Dir.}}^{c_1}}{\left| \det (D_{5w} - m_0)|_{\text{AP}}^{c_1 c_1^{-1}} \right|^{\frac{1}{2}}} e^{-i C_{5w}^{c_1}}, \tag{2.28}$$

where $D_0 = D|_{U=U_0}$, $i\eta_\mu^s(x) = \partial_s U_s(x,\mu) U_s(x,\mu)^{-1}$. The functional determinant of the domain wall fermion is defined in the interval $x_5 \in [-N+1, N]a_5$ of the extra fifth dimension (Shamir's version; Fig 1) with the Dirichlet b.c.(Dir.)

$$\psi_-(x, x_5)|_{x_5 = -Na_5} = 0, \qquad \psi_+(x, x_5)|_{x_5 = (N+1)a_5} = 0, \tag{2.29}$$

and that of the Pauli-Villars field is defined in $x_5 \in [-N+1, 3N]a_5$ with the anti-periodic one(AP). And $C_{5w}^{c_1}$ is a local counter term which can be constructed from the Chern-Simon current induced by the five-dimensional Wilson fermion.

In order to introduce chiral-asymmetric gauge interaction for the chiral zero modes at the two boundaries $x_5 = -(N+1)a_5$ and $Na_5$, the link field is assumed to be five-dimensional,

$$U(z,a) = \{U(x, x_5, \mu), U_5(x, x_5)\}, \quad z = (x, x_5), \quad a = 1, \ldots, 5, \tag{2.30}$$

interpolating between the four-dimensional link fields $U_0(x,\mu)$ and $U_1(x,\mu)$ at the boundaries (Figs 2 and 3). We assume that outside the finite interpolation region $x_5 \in [-\Delta, \Delta]$ ($\Delta < Na_5$) the link field does not depend on $x_5$ and $U_5(x, x_5) = 1$. We also assume that the five-dimensional gauge field $U(z,a)$ is smooth enough and satisfies the bound on the five-dimensional field strength as follows:

$$\|1 - P_{ab}(z)\| < \epsilon', \qquad \epsilon' < \frac{1}{50}. \tag{2.31}$$

Since the difference of the four-dimensional gauge fields at the boundaries is estimated as

$$\|U_0(x,\mu) - U_1(x,\mu)\| \simeq \frac{2\Delta}{a_5} \|1 - P_{5\mu}(z)\|, \tag{2.32}$$

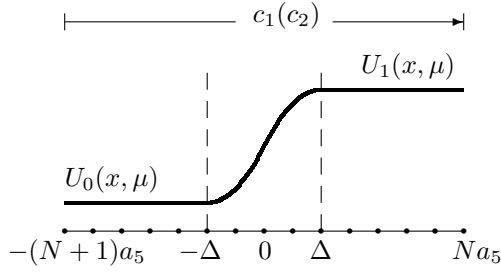

**Figure 2**. Domain wall fermion with a five-dimensional link field interpolating $U_0(x,\mu)$ and $U_1(x,\mu)$

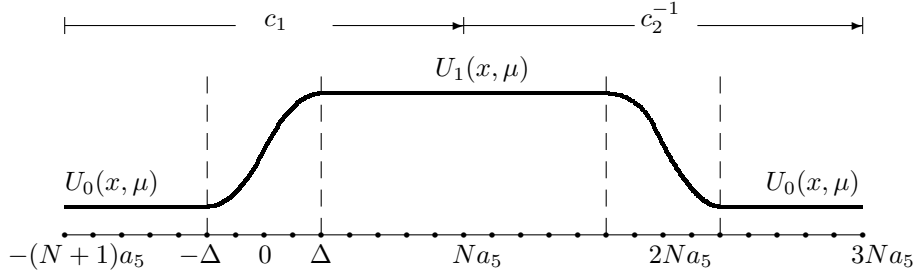

**Figure 3**. Pauli-Villars spinor-boson with a five-dimensional link field representing a closed loop $c_1 \, c_2^{-1}$ in the space of link fields

with the five dimensional bound on the field strength, we can make the interpolation smooth enough by taking $\Delta$ large enough. This condition assures that the Hamiltonian defined through the transfer matrix of the five-dimensional Wilson-Dirac fermion, $H = -\frac{1}{a_5}\ln T$, has a finite gap and makes the limit that the size of the fifth dimension goes to infinity, $N \to \infty$, well-defined.

Furthermore, this condition assures that the Chern-Simons current induced from the five-dimensional Wilson-Dirac fermion by the variation $\delta_\zeta U(z,a) = i\zeta_a(z) U(z,a)$,

$$\delta_\zeta \{\mathrm{Im}\det(D_{5\mathrm{w}} - m_0)|_{\mathrm{AP}}^{c_1 c_2^{-1}}\} = \sum_{x_5 \in c_1 c_2^{-1}} \sum_{x \in \Gamma} \zeta_a(z) j_a(z), \tag{2.33}$$

$$j_a(z) = \mathrm{Tr}\left(v_a(z) \frac{1}{D_{5\mathrm{w}} - m_0}\right)\bigg|_{\mathrm{AP}}, \tag{2.34}$$

$$v_a(z) = \left\{\frac{1}{2}(-\gamma_a - 1) U(z,a) \delta_{zz_1} \delta_{z_1+\hat{a},z_2} \right.$$
$$\left. +\frac{1}{2}(-\gamma_a + 1) U(z,a)^{-1} \delta_{zz_2} \delta_{z_1,z_2+\hat{a}}\right\}, \tag{2.35}$$

is a local functional of the link field. As shown in [21], when the anomaly cancellation condition is satisfied, $j_a(z)$ can be written with a local and gauge-invariant anti-symmetric tensor field $\chi_{ab}(z)$ as

$$j_a(z) = \partial_b^* \chi_{ba}(z), \qquad \chi_{ba}(z) = -\chi_{ab}(z), \tag{2.36}$$

and this tensor field $\chi_{ab}(z)$ gives the local counter term $C_{5\text{w}}^{c_1}$ as follows,

$$C_{5\text{w}}^{c_1} \equiv \sum_{x_5 \in c_1} \sum_{x \in \Gamma} \frac{1}{2} F_{ab}(z-\hat{b}) \chi_{ab}(z-\hat{b}). \tag{2.37}$$

This counter term cancels the bulk dependence of the five-dimensional interpolating link field $U(z, a)$ in the right-handed part of eq. (2.28), and makes it a functional of only the boundary values of the four-dimensional link fields $U_0(x, \mu), U_1(x, \mu)$.

## 3 Gradient flow for the admissible U(1) link fields

For the admissible U(1) link fields, the gradient flow can be formulated so that it preserves their topological charges. We adopt the following gradient-flow equation for the admissible U(1) link fields,

$$S(U_t) = 1/(4g_0^2) \sum_{x \in \Gamma} \sum_{\mu,\nu} F_{\mu\nu}(x,t)^2, \tag{3.1}$$

$$\frac{d}{dt} U_t(x, \mu) = -g_0^2 [\partial_{x,\mu} S(U_t)] U_t(x, \mu), \tag{3.2}$$

where the field tensor $F_{\mu\nu}(x, t)$ is defined through

$$F_{\mu\nu}(x, t) = i^{-1} \ln P_t(x, \mu, \nu), \qquad -\pi < F_{\mu\nu}(x, t) < \pi, \tag{3.3}$$

$$P_t(x, \mu, \nu) = U_t(x, \mu) U_t(x+\mu, \nu) U_t(x+\nu, \mu)^{-1} U_t(x, \nu)^{-1}, \tag{3.4}$$

and $\partial_{x,\mu}$ is the differential operator with respect to the U(1) link variable $U_t(x, \mu)$,

$$\partial_{x,\mu} f(U) = \left. \frac{d}{ds} f(U_s) \right|_{s=0}, \quad U_s(y, \nu) = \begin{cases} e^{is} U(x, \mu) & \text{if } (y, \nu) = (x, \mu), \\ U(y, \nu) & \text{otherwise}. \end{cases} \tag{3.5}$$

With this choice, the solution $U_t(x, \mu)$ preserves the admissibility condition,

$$|F_{\mu\nu}(x, t)| < \epsilon \quad \text{for all } x, \mu, \nu; \qquad \epsilon < \frac{1}{30}. \tag{3.6}$$

To see this fact, we first note that if we assume the Bianchi identity,

$$\partial_\mu F_{\nu\rho}(x, t) + \partial_\rho F_{\mu\nu}(x, t) + \partial_\nu F_{\rho\mu}(x, t) = 0, \tag{3.7}$$

then, by substituting the action (3.1) into the flow equation (3.2), it follows that

$$\frac{d}{dt} F_{\mu\nu}(x, t) = \partial_\rho^* \partial_\rho F_{\mu\nu}(x, t), \tag{3.8}$$

where $\partial_\rho$ and $\partial_\rho^*$ denote the forward and backward difference operators respectively. With this flow equation Eq. (3.8), the admissibility condition (3.6) is preserved by the following theorem (see Appendix B for the proof):

**Theorem 1 (maximum principle)** *Suppose $f(x,t)$ is periodic with respect to $x$, and satisfies the lattice version of the diffusion equation*

$$\frac{d}{dt} f(x,t) = \partial_\mu^* \partial_\mu f(x,t), \tag{3.9}$$

*where $x \in \Gamma$, $t \in [0,T]$. Then $f(x,t)$ has a maximum at $t=0$,*

$$\max_{x\in\Gamma,\ t\in[0,T]} f(x,t) = \max_{x\in\Gamma} f(x,0). \tag{3.10}$$

Conversely, if one is given the flow equation, Eq. (3.8), with the initial field tensor $F_{\mu\nu}(x,0)$ satisfying the admissibility condition Eq. (3.6) and the Bianchi identity Eq. (3.7) at $t=0$, then the solution $F_{\mu\nu}(x,t)$ respects the same condition and identity for $t>0$. From $F_{\mu\nu}(x,t)$, one can construct the transverse vector potential $A_\mu^T(x,t)$ by the relations (2.12), and then the link variables $U_t(x,\mu)$ by the relation (2.11). This $U_t(x,\mu)$ satisfies the original flow equation, Eqs. (3.2), (3.1).

It is straightforward to solve Eq. (3.8):

$$F_{\mu\nu}(x,t) = e^{-\Box t} F_{\mu\nu}(x), \qquad \Box = -\sum_\mu \partial_\nu^* \partial_\nu, \tag{3.11}$$

$$\tilde{F}_{\mu\nu}(p,t) = e^{-\hat{p}^2 t} \tilde{F}_{\mu\nu}(p), \qquad \hat{p}^2 = \sum_\nu [2\sin(p_\nu/2)]^2. \tag{3.12}$$

where the Fourier transformation of the field tensor is defined by

$$\tilde{F}_{\mu\nu}(p,t) = \sum_{x\in\Gamma} e^{-ip\cdot x} F_{\mu\nu}(x,t), \tag{3.13}$$

$$F_{\mu\nu}(x,t) = L^{-4} \sum_{x\in\Gamma} e^{ip\cdot x} \tilde{F}_{\mu\nu}(p,t). \tag{3.14}$$

Thus the transverse vector potential $A_\mu^T(x,t)$ is given by

$$A_\mu^T(x,t) = e^{-\Box t} A_\mu^T(x). \tag{3.15}$$

In the limit $t \to \infty$, only the zero mode $\tilde{F}_{\mu\nu}(0,t)$ remains, and we obtain

$$\lim_{t\to\infty} F_{\mu\nu}(x,t) = \tilde{F}_{\mu\nu}(0,0) = 2\pi m_{\mu\nu}/L^2, \tag{3.16}$$

where $m_{\mu\nu}$ is the magnetic flux of the initial link variable. Accordingly, we obtain

$$\lim_{t\to\infty} U_t(x,\mu) = U_{[w]}(x,\mu)\, \Lambda(x)\, \Lambda(x+\hat{\mu})^{-1}\, V_{[m]}(x,\mu) \equiv U_\star(x,\mu). \tag{3.17}$$

## 4 Proposal by Grabowska and Kaplan – Abelian case

We now implement the original proposal by Grabowska and Kaplan[1, 3] by using the domain wall fermion defined in the interval $x_5 \in [-N, N+1]$ of the extra fifth dimension

(Shamir's version). We can define the effective action in the formulation of Grabowska and Kaplan as

$$e^{\Gamma_{\rm GK}[U]} = \lim_{a_5 \to 0} \lim_{N \to \infty} \frac{\det (D_{5w} - m_0)|_{\rm Dir.}^{c_{\rm gf}}}{\left| \det (D_{5w} - m_0)|_{\rm AP}^{c_{\rm gf} c_{\rm gf}^{-1}} \right|^{\frac{1}{2}}} \quad (4.1)$$

$$= \det\{1 - P_+ + P_+ D Q_1 D_\star^\dagger\}, \quad (4.2)$$

where $D_\star = D|_{U=U_\star}$, and $c_{\rm gf}$ denotes the five dimensional link field $U(z,a)$ determined by the mapping

$$U(x, x_5, \mu) = U_t(x, \mu) \Big|_{t = -\frac{1}{p_0^2} \ln\left[\frac{\Delta + x_5}{2\Delta}\right]}, \quad x_5 \in [-\Delta, \Delta] \subset \mathbb{Z} a_5, \quad p_0 = \frac{2\pi}{L} \quad (4.3)$$

from the infinite gradient flow for the admissible U(1) link field, $U_t(x, \mu)$ ($t \in [0, \infty]$).

By comparing with (2.27), we immediately obtain the relation of $\Gamma_{\rm GK}[U]$ to the gauge-invariant effective action $\Gamma[U]$ as follows:

$$e^{\Gamma_{\rm GK}[U]} = e^{\Gamma[U]} e^{\Gamma[U_\star]^*} e^{+i \int_\infty^0 dt \, \mathcal{L}_{\eta_t}}, \quad (4.4)$$

or

$$\mathrm{Re}(\Gamma_{\rm GK}[U]) = \mathrm{Re}(\Gamma[U] + \Gamma[U_\star]) \quad (4.5)$$

$$= \frac{1}{2} \mathrm{Tr} \, \mathrm{Ln} \, D + \frac{1}{2} \mathrm{Tr} \, \mathrm{Ln} \, D_\star, \quad (4.6)$$

$$\mathrm{Im}(\Gamma_{\rm GK}[U]) = \mathrm{Im}(\Gamma[U] - \Gamma[U_\star]) + \int_\infty^0 dt \, \mathcal{L}_{\eta_t}. \quad (4.7)$$

Here the variational parameter $\eta_\mu(x, t)$ for the gradient flow is given by

$$\eta_\mu(x, t) = -i \, \partial_t U_t(x, \mu) U_t(x, \mu)^{-1}$$

$$= \partial_\nu^* F_{\nu\mu}(x, t) \quad (4.8)$$

$$= e^{-\Box t} \partial_\nu^* F_{\nu\mu}(x), \quad (4.9)$$

and the measure term $\mathcal{L}_{\eta_t}$ is given by

$$\mathcal{L}_{\eta_t} = \sum_{x \in \Gamma} (e^{-\Box t} \partial_\nu^* F_{\nu\mu}(x)) \, j_\mu(x)|_{U=U_t}. \quad (4.10)$$

## 5 Locality

In this section, we examine the locality property of the measure term contribution in $\mathrm{Im}(\Gamma_{\rm GK})$. For this purpose, we consider the topologically trivial sector in the infinite volume limit.

The measure term current $j_\mu(x)|_{U=U_t}$ is a gauge-invariant local field of $U_t(x,\mu)$. It can be expanded perturbatively in the power series of the flowed field tensor $F_{\mu\nu}(x,t) = e^{-\Box t} F_{\mu\nu}(x)$. In the momentum space, it is given as

$$j_\mu(p) = \sum_{k \geq 4} \int V^{(k)}_{\mu\mu_1\nu_1\cdots\mu_k\nu_k}(p, p_1, \ldots, p_{k-1})(2\pi)^4 \delta^{(4)}\Big(\sum_{i=1}^k p_i\Big) \prod_{i=1}^k e^{-\hat{p}_i^2 t} \tilde{F}_{\mu_i\nu_i}(p_i) \frac{d^4 p_i}{(2\pi)^4}, \quad (5.1)$$

where the Fourier transformation is defined through

$$\tilde{A}_\mu(p) = \sum_{x \in \mathbb{Z}^4} e^{ip\cdot x} A_\mu(x), \qquad p_\mu \in [0, 2\pi), \quad (5.2)$$

$$A_\mu(x) = \int \frac{d^4 p}{(2\pi)^4} e^{-ip\cdot x} \tilde{A}_\mu(p), \quad (5.3)$$

and $\hat{p}_i^2 = \sum_\mu [2\sin((p_i)_\mu/2)]^2$. $V^{(k)}_{\mu\mu_1\nu_1\cdots\mu_k\nu_k}(p, p_1, \ldots, p_{k-1})$ are analytic functions with respect to the momenta $p, p_1, \ldots p_{k-1}$. We note that perturbation series of the current start from the quartic term because of the anomaly cancellation condition and the lattice symmetries.[5]

Then, substituting Eq. (5.1) to the measure term (4.10) and integrating with respect to the flow time $t$, we obtain the following expression for the measure term contribution.

$$-\int_\infty^0 dt\, \mathcal{L}_{\eta_t} = \sum_{k \geq 5} \int \frac{V'^{(k)}_{\mu_1\cdots\mu_k}(p_1, \ldots, p_{k-1})}{\sum_{i=1}^k \hat{p}_i^2} (2\pi)^4 \delta^{(4)}\Big(\sum_{i=1}^k p_i\Big) \prod_{i=1}^k \tilde{A}_{\mu_i}(p_i) \frac{d^4 p_i}{(2\pi)^4}$$

$$\equiv \sum_{k \geq 5} \int \tilde{\Gamma}^{(k)}_{\mu_1\cdots\mu_k}(p_1, \ldots, p_{k-1})(2\pi)^4 \delta^{(4)}\Big(\sum_{i=1}^k p_i\Big) \prod_{i=1}^k \tilde{A}_{\mu_i}(p_i) \frac{d^4 p_i}{(2\pi)^4}. \quad (5.4)$$

Here $V'^{(k)}_{\mu_1\cdots\mu_k}(p_1, \ldots, p_k)$ are the products of analytic functions $V^{(k-1)}_{\mu\mu_1\nu_1\cdots\mu_{k-1}\nu_{k-1}}(p, p_1, \ldots, p_{k-1})$ and $(k+2)$-th degree polynomials of the momenta $(\hat{p}_i)_\mu = 2\sin((p_i)_\mu/2)$ coming from the field tensor $\tilde{F}_{\mu\nu}(p_i) = (\hat{p}_i)_\mu \tilde{A}_\nu(p_i) - (\hat{p}_i)_\nu \tilde{A}_\mu(p_i)$ and its derivatives in the measure term (4.10).[6] The factor $1/\sum_{i=1}^k \hat{p}_i^2$ comes from the infinite gradient-flow.

From this result, we note that $\tilde{\Gamma}^{(k)}_{\mu_1\cdots\mu_k}(p_1, \ldots, p_k)$ are $(k-1)$-times differentiable functions with respect to all the variables $(p_i)_\mu \in [0, 2\pi)$. $k$-th derivatives of $\tilde{\Gamma}^{(k)}_{\mu_1\cdots\mu_k}(p_1, \ldots, p_{k-1})$ are smooth and bounded when $\sum_i \hat{p}_i^2 \neq 0$. However, when $\sum_i \hat{p}_i^2 = 0$, $k$-th derivatives are singular as far as the accidental cancellation of $V'^{(k)}_{\mu_1\cdots\mu_k}(p_1, \ldots, p_{k-1})$ does not happen. The

---

[5]In two-dimensions, it starts from the cubic term.

[6] At least two derivatives are necessary so that $j_\mu(x)$ transforms as an axial vector under the lattice symmetries. For example,

$$\epsilon_{\alpha\beta\gamma\delta} F_{\alpha\beta} F_{\gamma\delta} \partial^*_\mu F_{\mu\nu} \partial^*_\nu F_{\rho\sigma} F_{\rho\sigma} \quad (5.5)$$

is a possible term in the measure term $\mathcal{L}_\eta$.

Fourier transformation of $k$-th derivatives of $\tilde{\Gamma}^{(k)}_{\mu_1\cdots\mu_k}(p_1,\ldots,p_{k-1})$ are given by

$$(i\partial_{(p_i)_\mu})^\alpha \tilde{\Gamma}^{(k)}_{\mu_1\cdots\mu_k}(p_1,\ldots,p_{k-1})$$
$$= \sum_{x_1,\ldots,x_{k-1}\in\mathbb{Z}^4} e^{-(ip_1\cdot x_1+\cdots+ip_{k-1}\cdot x_{k-1})}(x_i)^\alpha_\mu \Gamma^{(k)}_{\mu_1\cdots\mu_k}(x_1,\ldots,x_{k-1},0)$$
$$= \sum_n \sum_{B_n^{(k)}} e^{-(ip_1\cdot x_1+\cdots+ip_{k-1}\cdot x_{k-1})}(x_i)^\alpha_\mu \Gamma^{(k)}_{\mu_1\cdots\mu_k}(x_1,\ldots,x_{k-1},0), \quad (5.6)$$

where we take $x_k$ as the reference point, and $B_n^{(k)}$ is defined by

$$B_n^{(k)} = \left\{(x_1,\ldots,x_{k-1}) \in \mathbb{Z}^4 \times \cdots \times \mathbb{Z}^4 \mid ||x_1|| + \cdots + ||x_{k-1}|| = n\right\}. \quad (5.7)$$

We note that the number of elements in a set $B_n^{(k)}$ is estimated as follows:

$$\sum_{i=1}^{4k-4} 2^i \binom{4k-4}{i}\binom{n-1}{i-1} = O(n^{4k-5}). \quad (5.8)$$

When $\alpha \geq k$, the right-hand side of Eq. (5.6) is not absolutely convergent. That is,

$$n^{4k-5} n^\alpha \Gamma^{(k)}_{\mu_1\cdots\mu_k}(x_1,\ldots,x_{k-1},0) \quad (5.9)$$

are not absolutely convergent series. Therefore for every positive integer $N$, there exist $n > N$ and $x_1,\ldots,x_{k-1}$ such that

$$\left|\Gamma^{(k)}_{\mu_1\cdots\mu_k}(x_1,\ldots,x_{k-1},0)\right| > \frac{c}{n^{5k-4}}, \quad (5.10)$$

where $c$ is a positive constant independent of $n$ and $x_1,\ldots,x_{k-1}$. (In the two-dimensional case, the corresponding result reads $> c/n^{3k-2}$.) This inequality shows that the measure term contribution in the imaginary part of effective action $\text{Im}(\Gamma_{\text{GK}}[U])$ is not exponentially bounded and thus non-local.

We can also derive an upper bound of $\Gamma^{(k)}_{\mu_1\cdots\mu_k}(x_1,\ldots,x_{k-1},0)$ in the following manner. Starting from the Fourier transformation,

$$\Gamma^{(k)}_{\mu_1\cdots\mu_k}(x_1,\ldots,x_{k-1},0)$$
$$= \int \frac{d^4p_1}{(2\pi)^4}\cdots\frac{d^4p_{k-1}}{(2\pi)^4} e^{ip_1\cdot x_1+\cdots+ip_{k-1}\cdot x_{k-1}}\tilde{\Gamma}^{(k)}_{\mu_1\cdots\mu_k}(p_1,\ldots,p_{k-1}), \quad (5.11)$$

we have

$$(x_i)^\alpha_\mu \Gamma^{(k)}_{\mu_1\cdots\mu_k}(x_1,\ldots,x_{k-1},0)$$
$$= \int \frac{d^4p_1}{(2\pi)^4}\cdots\frac{d^4p_{k-1}}{(2\pi)^4} e^{ip_1\cdot x_1+\cdots+ip_{k-1}\cdot x_{k-1}}(i\partial_{(p_i)_\mu})^\alpha \tilde{\Gamma}'^{(k)}_{\mu_1\cdots\mu_k}(p_1,\ldots,p_{k-1}), \quad (5.12)$$

where $\alpha \leq k$. We note that

$$\int_a^b dp\, e^{ipx} f(p) = O(1/x), \quad (5.13)$$

where $f(p)$ is a smooth function on $[a,b]$. Since $k$-th derivatives of $\tilde{\Gamma}'^{(k)}_{\mu_1\cdots\mu_k}(p_1,\ldots,p_{k-1})$ are smooth and bounded from above and below when $\sum_i \hat{p}_i^2 \neq 0$, the left-hand side of Eq. (5.12) converges as $(x_i)_\mu$ approaches infinity,

$$(x_i)^\alpha_\mu \Gamma^{(k)}_{\mu_1\cdots\mu_k}(x_1,\ldots,x_{k-1},0) \to 0. \tag{5.14}$$

Thus we have the following upper bound,

$$\Gamma^{(k)}_{\mu_1\cdots\mu_k}(x_1,\ldots,x_{k-1},0) < \frac{c'}{||x_i||^k}, \tag{5.15}$$

where $c'$ is a positive constant independent of $x_1,\ldots,x_{k-1}$.

## 6 Discussion

We examined the proposal by Grabowska and Kaplan (GK) to use the infinite gradient flow in the domain-wall formulation of chiral lattice gauge theories. We considered the case of Abelian theories and formulated the gradient flow for the admissible U(1) link fields so that it preserves their topological charges. The GK's effective action turns out to be equal to the sum of the Lüscher's gauge-invariant effective actions for the target Weyl fermions and the mirror "fluffy" fermions, plus the so-called measure-term integrated along the infinite gradient flow. The measure-term current is originally a local(analytic) and gauge-invariant functional of the admissible link field. However, with the infinite gradient flow, it gives rise to non-local(non-analytic) vertex functions which are not suppressed exponentially at large distance. The "fluffy" fermions remains as a source of non-local contribution, which couple yet to the Wilson-line and magnetic-flux degrees of freedom of the dynamical link field.

The possible effect of the non-local terms which appear in the imaginary part of the effective action, $\text{Im}(\Gamma_{\text{GK}}[U])$, should be examined further. In particular, there may be some effect on the critical behavior and the continuum limit of the lattice model (cf. [23, 24]). In this respect, we note that the vertex functions are non-local due to the factor $1/\left[\sum_{i=1}^k \hat{p}_i^2\right]$. They are singular only when all momenta vanish. Then it is not likely that these vertex functions affect the Wilsonian renormalization group and therefore the critical behavior of the lattice model. On the other hand, the original singularity, which appears at the point where all momenta vanish, remains in IR. This question is a dynamical one and would require non-perturbative and/or numerical approaches. We believe, however, it should be addressed before considering the use of the (infinite) gradient flow method in the construction of chiral lattice gauge theories (cf. [25, 26]).

## A  More on the chiral solution of the GW relation

In this appendix, we extend the derivation of the chiral solution of the Ginsparg-Wilson relation, $D_\chi$, for the case of sudden flow[2] to the case of gradual flow, and show that the result of the latter is identical to the former.

In [2], the chiral solution of the Ginsparg-Wilson relation is obtained by the formula,

$$D_\chi = \frac{1}{2}(1 + \gamma_5 \mathcal{E}_\chi), \tag{A.1}$$

where

$$\mathcal{E}_\chi = \lim_{L \to \infty} \mathcal{E}_\chi^{(L)}, \qquad \mathcal{E}_\chi^{(L)} = \frac{1 - \prod_{s=L}^{1} T(s)}{1 + \prod_{s=L}^{1} T(s)}. \tag{A.2}$$

Here the set of the link fields $\{U^{(s)}(x,\mu) \,|\, s = 1, \cdots, L\}$ defines the extrapolation of the dynamical four-dimentional link variables $U(x,\mu)$ located at $s = 1$ by the gradient flow: $U^{(1)}(x,\mu) = U(x,\mu)$. The link field at $s = L$ is denoted as $U_\diamond(x,\mu)$ in general, while the fixed point of the gradient flow is denoted with the symbol $\star$ as $U_\star(x,\mu)$. We note that $U_\diamond(x,\mu)$ is not necessarily equal to $U_\star(x,\mu)$, when the region of the extrapolation has a finite extent (in lattice unit). We also note a useful matrix relation,

$$\frac{2AB}{1 + AB} = \frac{2A}{1 + A} \frac{1}{1 + \frac{1-B}{1+B}\frac{1-A}{1+A}} \frac{2B}{1 + B}. \tag{A.3}$$

### A.1 Sudden flow (to the fixed point)

First we assume

$$\prod_{s=L}^{1} T(s) = T_\diamond^{L/2} T^{L/2}, \tag{A.4}$$

and introduce

$$\epsilon_\diamond^{(L)} = \frac{1 - T_\diamond^{L/2}}{1 + T_\diamond^{L/2}}, \qquad \epsilon^{(L)} = \frac{1 - T^{L/2}}{1 + T^{L/2}}. \tag{A.5}$$

Set

$$A = T_\diamond^{L/2}, \qquad B = T^{L/2}, \tag{A.6}$$

and use the matrix relation (A.3). Then we obtain

$$1 - \mathcal{E}_\chi^{(L)} = (1 - \epsilon_\diamond^{(L)}) \frac{1}{1 + \epsilon^{(L)} \epsilon_\diamond^{(L)}} (1 - \epsilon^{(L)}). \tag{A.7}$$

Now we take the limit $L \to \infty$ and obtain

$$1 - \mathcal{E}_\chi = (1 - \epsilon_\diamond) \frac{1}{1 + \epsilon \epsilon_\diamond} (1 - \epsilon). \tag{A.8}$$

In this result, we can replace $U_\diamond$ with $U_\star$ and reproduce the original result [2],

$$1 - \mathcal{E}_\chi = (1 - \epsilon_\star) \frac{1}{1 + \epsilon \epsilon_\star} (1 - \epsilon). \tag{A.9}$$

### A.2 Gradual flow in the finite extent $\Delta$

Next we assume

$$\prod_{s=\Delta+L/2+1}^{-L/2+1} T(s) = T_\diamond^{L/2} \times \prod_{s=\Delta}^{1} T(s) \times T^{L/2}. \tag{A.10}$$

Set
$$A = T_\diamond^{L/2}, \qquad B = \prod_{s=\Delta}^{1} T(s), \qquad C = T^{L/2}, \tag{A.11}$$

and use the matrix relation (A.3) twice. Then we obtain

$$\begin{aligned}
1 - \mathcal{E}_\chi^{(L)} &= (1-\epsilon_\diamond^{(L)}) \frac{1}{1 + \frac{1-BC}{1+BC}\epsilon_\diamond^{(L)}} \frac{2BC}{1+BC} \\
&= (1-\epsilon_\diamond^{(L)}) \frac{1}{1 + \left(1 - \frac{2BC}{1+BC}\right)\epsilon_\diamond^{(L)}} \frac{2BC}{1+BC} \\
&= (1-\epsilon_\diamond^{(L)}) \frac{1}{(1+\epsilon_\diamond^{(L)}) - \frac{2BC}{1+BC}\epsilon_\diamond^{(L)}} \frac{2BC}{1+BC},
\end{aligned} \tag{A.12}$$

and

$$\begin{aligned}
\frac{2BC}{1+BC} &= \frac{2B}{1+B} \frac{1}{1 + \epsilon^{(L)}\frac{1-B}{1+B}}(1-\epsilon^{(L)}) \\
&= 2B \frac{1}{(1+B) + \epsilon^{(L)}(1-B)}(1-\epsilon^{(L)}) \\
&= 2B \frac{1}{(1+\epsilon^{(L)}) + (1-\epsilon^{(L)})B}(1-\epsilon^{(L)}) \\
&= \frac{2}{(1+\epsilon^{(L)})B^{-1} + (1-\epsilon^{(L)})}(1-\epsilon^{(L)}).
\end{aligned} \tag{A.13}$$

Combining these two results, we further obtain

$$\begin{aligned}
1 - \mathcal{E}_\chi^{(L)} &= (1-\epsilon_\diamond^{(L)}) \frac{1}{(1+\epsilon_\diamond^{(L)}) - \left(\frac{2}{(1+\epsilon^{(L)})B^{-1}+(1-\epsilon^{(L)})}(1-\epsilon^{(L)})\right)\epsilon_\diamond^{(L)}} \times \\
&\quad \frac{2}{(1+\epsilon^{(L)})B^{-1} + (1-\epsilon^{(L)})}(1-\epsilon^{(L)}).
\end{aligned} \tag{A.14}$$

By canceling the same matrix factor,

$$\frac{2}{(1+\epsilon^{(L)})B^{-1} + (1-\epsilon^{(L)})}, \tag{A.15}$$

in the numerator and the denominator, we finally obtain

$$\begin{aligned}
1 - \mathcal{E}_\chi^{(L)} &= (1-\epsilon_\diamond^{(L)}) \frac{2}{(1+\epsilon^{(L)})B^{-1}(1+\epsilon_\diamond^{(L)}) + (1-\epsilon^{(L)})(1-\epsilon_\diamond^{(L)})}(1-\epsilon^{(L)}) \\
&= (1-\epsilon_\diamond^{(L)}) \frac{2}{(1+\epsilon^{(L)})\left(\prod_{s=\Delta}^1 T(s)\right)^{-1}(1+\epsilon_\diamond^{(L)}) + (1-\epsilon^{(L)})(1-\epsilon_\diamond^{(L)})}(1-\epsilon^{(L)}).
\end{aligned} \tag{A.16}$$

Now we take the limit $L \to \infty$ with $\Delta$ kept finite and obtain

$$\begin{aligned}
1 - \mathcal{E}_\chi &= (1-\epsilon_\diamond) \frac{2}{(1+\epsilon)\left(\prod_{s=\Delta}^1 T(s)\right)^{-1}(1+\epsilon_\diamond) + (1-\epsilon)(1-\epsilon_\diamond)}(1-\epsilon) \\
&= 2\hat{P}_{-\diamond} \frac{1}{\hat{P}_+\left(\prod_{s=\Delta}^1 T(s)\right)^{-1}\hat{P}_{+\diamond} + \hat{P}_-\hat{P}_{-\diamond}} \hat{P}_-.
\end{aligned} \tag{A.17}$$

In the above result, if one sets $\Delta = 0$ or replaces the operator $\prod_{s=\Delta}^{1} T(s)$ to unity, it reproduces the result of the case of sudden flow, as it should be.

## A.3 Equivalence of the two cases

The results of these two cases are actually identical. Let us introduce the chiral basis $u$, $v$ of $\hat{P}_{\pm}$, and the chiral basis $u_\diamond$, $v_\diamond$ of $\hat{P}\pm\diamond$. Then the identity operator can be expressed as

$$1 = uu^\dagger + vv^\dagger = u_\diamond u_\diamond^\dagger + v_\diamond v_\diamond^\dagger. \tag{A.18}$$

With these bases, the operator in the denominator is expressed as

$$\left[\hat{P}_+ \big(\prod_{s=\Delta}^{1} T(s)\big)^{-1} \hat{P}_{+\diamond} + \hat{P}_- \hat{P}_{-\diamond}\right]$$
$$= (uu^\dagger + vv^\dagger) \left[\hat{P}_+ \big(\prod_{s=\Delta}^{1} T(s)\big)^{-1} \hat{P}_{+\diamond} + \hat{P}_- \hat{P}_{-\diamond}\right] (u_\diamond u_\diamond^\dagger + v_\diamond v_\diamond^\dagger)$$
$$= \left[u\{u^\dagger \big(\prod_{s=\Delta}^{1} T(s)\big)^{-1} u_\diamond\} u_\diamond^\dagger + v\{v^\dagger v_\diamond\} v_\diamond^\dagger\right]. \tag{A.19}$$

Its inverse therefore can be expressed as

$$\left[\hat{P}_+ \big(\prod_{s=\Delta}^{1} T(s)\big)^{-1} \hat{P}_{+\diamond} + \hat{P}_- \hat{P}_{-\diamond}\right]^{-1}$$
$$= \left[u_\diamond \{u^\dagger \big(\prod_{s=\Delta}^{1} T(s)\big)^{-1} u_\diamond\}^{-1} u^\dagger + v_\diamond \{v^\dagger v_\diamond\}^{-1} v^\dagger\right]. \tag{A.20}$$

Using this, one obtains

$$1 - \mathcal{E}_\chi = 2\, v_\diamond \{v^\dagger v_\diamond\}^{-1} v^\dagger. \tag{A.21}$$

Thus, the result for gradual flow does not actually depend on the operator $\prod_{s=\Delta}^{1} T(s)$. Therefore it is identical to that of the case of sudden flow, where the operator is set to unity.

Since the result of the case of gradual flow is independent of $\prod_{s=\Delta}^{1} T(s)$, one can take the limit $\Delta \to \infty$ or $U_\diamond \to U_\star$, and obtain the original result

$$1 - \mathcal{E}_\chi = (1 - \epsilon_\star) \frac{1}{1 + \epsilon \epsilon_\star} (1 - \epsilon) \tag{A.22}$$
$$= 2\, v_\star \{v^\dagger v_\star\}^{-1} v^\dagger. \tag{A.23}$$

Thus it is indeed the general result for $\mathcal{E}_\chi$ which is valid not only for the sudden flow, but also for the gradual flow.

## B   Proof of maximum principle

To prove Theorem 1 we first note that

$$g(x,t) \equiv f(x,t) - \epsilon t, \quad \epsilon > 0 \tag{B.1}$$

satisfies the inequality

$$\frac{d}{dt}g(x,t) < \partial_\mu^* \partial_\mu g(x,t). \tag{B.2}$$

Let $g(x,t)$ has a maximum at $(x_0, t_0)$ with $t_0 > 0$. Then

$$\frac{d}{dt}g(x_0, t_0) > 0, \qquad \partial_\mu^* \partial_\mu g(x_0, t_0) \leq 0. \tag{B.3}$$

However these inequalities contradict the inequality (B.2). Therefore

$$\max_{x \in \Gamma, \ t \in [0,T]} g(x,t) = \max_{x \in \Gamma} g(x,0). \tag{B.4}$$

Finally taking the limit $\epsilon \to +0$, we have Eq. (3.10).

## References

[1] D. M. Grabowska and D. B. Kaplan, Phys. Rev. Lett. **116**, no. 21, 211602 (2016) doi:10.1103/PhysRevLett.116.211602 [arXiv:1511.03649 [hep-lat]].

[2] D. M. Grabowska and D. B. Kaplan, Phys. Rev. D **94**, no. 11, 114504 (2016) doi:10.1103/PhysRevD.94.114504 [arXiv:1610.02151 [hep-lat]].

[3] D. B. Kaplan and D. M. Grabowska, PoS LATTICE **2016**, 018 (2016). doi:10.22323/1.256.0018

[4] D. B. Kaplan, Phys. Lett. B **288**, 342 (1992) [arXiv:hep-lat/9206013].

[5] Y. Shamir, Nucl. Phys. B **406**, 90 (1993) [arXiv:hep-lat/9303005].

[6] R. Narayanan and H. Neuberger, JHEP **0603**, 064 (2006) doi:10.1088/1126-6708/2006/03/064 [hep-th/0601210].

[7] M. Lüscher, JHEP **1008**, 071 (2010) Erratum: [JHEP **1403**, 092 (2014)] doi:10.1007/JHEP08(2010)071, 10.1007/JHEP03(2014)092 [arXiv:1006.4518 [hep-lat]].

[8] M. Lüscher and P. Weisz, JHEP **1102**, 051 (2011) doi:10.1007/JHEP02(2011)051 [arXiv:1101.0963 [hep-th]].

[9] M. Lüscher, PoS LATTICE **2013**, 016 (2014) doi:10.22323/1.187.0016 [arXiv:1308.5598 [hep-lat]].

[10] K. i. Okumura and H. Suzuki, PTEP **2016**, no. 12, 123B07 (2016) doi:10.1093/ptep/ptw167 [arXiv:1608.02217 [hep-lat]].

[11] H. Makino and O. Morikawa, PTEP **2016**, no. 12, 123B06 (2016) doi:10.1093/ptep/ptw183 [arXiv:1609.08376 [hep-lat]].

[12] H. Makino, O. Morikawa and H. Suzuki, PTEP **2017**, no. 6, 063B08 (2017) doi:10.1093/ptep/ptx085 [arXiv:1704.04862 [hep-lat]].

[13] M. Lüscher, Nucl. Phys. B **549**, 295 (1999) [arXiv:hep-lat/9811032].

[14] P. H. Ginsparg and K. G. Wilson, Phys. Rev. D **25**, 2649 (1982).

[15] M. Lüscher, Phys. Lett. B **428**, 342 (1998) [arXiv:hep-lat/9802011].

[16] H. Neuberger, Phys. Lett. B **417**, 141 (1998) [arXiv:hep-lat/9707022].

[17] H. Neuberger, Phys. Lett. B **427**, 353 (1998) [arXiv:hep-lat/9801031].

[18] R. Narayanan and H. Neuberger, Nucl. Phys. B **412**, 574 (1994) [arXiv:hep-lat/9307006].

[19] R. Narayanan and H. Neuberger, Phys. Rev. Lett. **71**, 3251 (1993) [arXiv:hep-lat/9308011].

[20] R. Narayanan and H. Neuberger, Nucl. Phys. B **443**, 305 (1995) [arXiv:hep-th/9411108].

[21] Y. Kikukawa, Phys. Rev. D **65**, 074504 (2002) [arXiv:hep-lat/0105032].

[22] P. Hernandez, K. Jansen and M. Lüscher, Nucl. Phys. B **552**, 363 (1999) [arXiv:hep-lat/9808010].

[23] J. Sak, Phys. Rev. B **8**, 281 (1973).

[24] M. Suzuki, Y. Yamazaki and G. Igarashi, Phys. Lett. **42A**, 313 (1972).

[25] H. Fukaya, T. Onogi, S. Yamamoto and R. Yamamura, PTEP **2017**, no. 3, 033B06 (2017) doi:10.1093/ptep/ptx017 [arXiv:1607.06174 [hep-th]].

[26] Y. Hamada and H. Kawai, PTEP **2017**, no. 6, 063B09 (2017) doi:10.1093/ptep/ptx086 [arXiv:1705.01317 [hep-lat]].